# THE SYMMETRY OF THE GENETIC CODE AND A UNIVERSAL TREND OF AMINO ACID GAIN AND LOSS

Semenov D.A. (dasem@mail.ru)

International Research Center for Studies of Extreme States of the Organism at the Presidium of the Krasnoyarsk Research Center, Siberian Branch of the Russian Academy of Sciences

Part 1 of the study intends to show that the universal trend of amino acid gain and loss discovered by Jordan et al. (2005) can be accounted for by the spontaneity of DNA typical damages. These damages lead to replacements of guanine and cytosine by thymine. Part 2 proposes a hypothesis of the evolution of the genetic code, the leading mechanism of which is the nucleotide spontaneous damage. The hypothesis accounts for the universal trend of amino acid gain and loss, stability of the genetic code towards point mutations, the presence of code dialects, and the symmetry of the genetic code table.

**Part 1.** In 2005, Nature published a study (Jordan et al., 2005) that demonstrated that such amino acids as His, Cys, Phe, Met, and Ser accrue in all proteins of different organisms whereas Pro, Ala, Glu, and Gly are consistently lost. The authors related their results to the evolution of the genetic code, referring to Trifonov (2004). The discovery caused some controversy among scientists. Zuckerkandl, a recognized authority in molecular biology informed Jordan and colleagues that he had proposed a relationship between such replacements and the evolution of the genetic code thirty years before (Zuckerkandl et al., 1971). Reputed specialists in the evolution of the genetic code tried to prove that the discovered trend was an artifact (Hurst et al., 2006).

My goal is to convince the reader that there can be a much simpler explanation of this trend. Let us first consider the symmetry of the genetic code. Researchers often relate the symmetry of the genetic code to its origin. A pioneer study in this field was performed by Hornos and colleagues (1993). In my study, the function of the symmetry will be to help us illustrate the main thesis suggesting that the origin of the trend of amino acid gain and loss is related to the damage of DNA. To begin with, we should refer to the very first study of the symmetry of the genetic code.

|   | C | G | U | A |
|---|---|---|---|---|
| C | CC | CG | CU | CA |
| G | GC | GG | GU | GA |
| U | UC | UG | UU | UA |
| A | AC | AG | AU | AA |

Table 1. A table of base pairs (called "roots" by Yu.B. Rumer (1966, 1968, 1969)) of the genetic code. "Strong" base pairs are indicated in red.

The first study of the structure of the genetic code was reported by Yu.B. Rumer in 1966 (Rumer, 1966). In that study and in the subsequent ones (1968, 1969) Rumer analyzed the symmetry of the table of the universal genetic code. The author focused his attention on the presence of the base pairs (i.e. the first two nucleotides of the triplet) and their ability or inability to encode just one amino acid. Of 16 base pairs, 8 were strong (encoding just one amino acid) and 8 were weak (encoding more than one amino acid). Having analyzed the table of the universal genetic code, Rumer proposed the following "strength"-based sequence of nucleotides: C>G>U>A (Table 1). He based his canonical sequence of nucleotides on their ability to form three or two hydrogen bonds with complementary bases in the DNA double helix.

Note that "strong" base pairs are lost and "weak" gained (Table 2). Of the four strongest base pairs (CC, GC, CG, and GG) three are lost. Whether CG is lost remains unclear. This base pair encodes arginine, which is also encoded by AG, a weak base pair. On average, arginine accumulates, but it is not clear which base pair is responsible for this. One can say that the universal trend of amino acid gain makes base pairs AT-richer.

|   | C  | G  | U  | A  |
|---|----|----|----|----|
| C | CC | CG | CU | CA |
| G | GC | GG | GU | GA |
| U | UC | UG | UU | UA |
| A | AC | AG | AU | AA |

Table 2. The distribution of losers (blue) and gainers (red) in the table of genetic code base pairs.

Let us examine the amino acid serine. It is encoded by two types of codons. To demonstrate the general thesis about the gain of weak base pairs, it would be good if AG were a gainer. Note that consistent loss of glutamine (Glu) also contradicts the proposed principle. Let us suppose that the UC base pair is a gainer. Then note that GC, GG, and GA are lost and UC, UG, and UU gained. This may be indicative of the significance of the oxidative guanine damage (G→T). This mechanism may make a considerable contribution to the trend, such transitions being rather harmless.

Rumer related the symmetry of the genetic code to the complementarity of the A-T and G-C nucleotides. Note that this complementarity is at least twofold. First, this is the complementarity of the nucleotides in the DNA double helix. Second, mRNA's codon is complementary to tRNA's anticodon. Moreover, in the latter case, it would be correct to speak of base pair complementarity. This idea is illustrated well by the mitochondrial genetic code, in which just 22 tRNAs are used.

There is a hypothesis proposing that RNA existed before DNA did. DNA emerged as another structure featuring complementarity and Rumer's symmetry. **Thus, the very emergence of DNA, whenever it occurred, should have affected the proportions of amino acids in proteins.** Below is my attempt to prove this.

There is an opinion that thymine was incorporated in DNA to prevent cytosine from being replaced by uracil in the course of deamination. This assumption, however, seems groundless. Incorporation of thymine actually triggered new mutations. Investigation of the two abovementioned characteristic DNA damages, cytosine deamination and oxidative guanine damage, will definitely point to the source causing deviation from detailed equilibrium in the course of amino acid replacements. The C→T and G→T damages are certainly repaired, but there is always a nonzero probability that repair will not occur before replication starts. Thus, DNA continuously gains AT-pairs. The occurrence of thymine in DNA is almost as likely as the occurrence of uracil in RNA. About 50% of cytosine in DNA is methylated and deamination converts cytosine to thymine. Oxidative guanine damage leads to the replacement of guanine by thymine only due to the specific structure of the DNA double helix. Uracil must be less helpful in maintaining the double helix structure, so nature chose to enhance information security dramatically and pay for this with systematic accumulation of thymine. The reverse process – replacement of AT-pairs by GC – is less spontaneous.

Of the greatest interest is aimlessness of amino acid replacements. All gained amino aids are perceived as significant, probably because they are rare. However, the reason for their gain may be not that they are necessary but that their accumulation is relatively harmless. I should note here that there is a deviation from detailed equilibrium in the very nature of amino acid replacements: if a rare amino acid is lost, it has most probably been significant and this replacement is punished by natural selection; if a rare amino acid replaces another one, less rare, it does not necessarily perform an important function.

Table 3 is a schematic representation of all possible amino acid replacements that result from nucleotide replacements in codon base pairs. The C→A and G→A replacements are equivalent to the G→T and C→T replacements in DNA complementary nucleotides, respectively. As a first approximation, let us prohibit replacements that would change the polarity of amino acids; then all possible replacements can be presented as four graphs beginning at the four strongest base pairs and three short graphs (Table 3). Crosses denote impossible routes for extending the graph, which would cause a change in the polarity of the amino acid. No duplicate replacements are shown.

| AU | ← | CU | → | UU | | AU | | CU | | UU |
|---|---|---|---|---|---|---|---|---|---|---|
| × | | ↑ | | × | | × | | × | | |
| AC | × | CC | × | UC | | AG | ← | CG | × | UG |
| ↓ | | × | | ↓ | | ↓ | | ↓ | | |
| AA | | CA | × | UA | | AA | ← | CA | | UA |

| AU | ← | GU | → | UU | | AU | ← | GU | → | UU |
|---|---|---|---|---|---|---|---|---|---|---|
| ↑ | | ↑ | | ↑ | | | | ↑ | | × |
| AG | ← | GG | → | UG | | AC | × | GC | × | UC |
| | | × | | × | | | | × | | |
| AA | | GA | | UA | | AA | ← | GA | → | UA |

Table 3. Possible replacements due to AT-enrichment, with the polarity of amino acids remaining unchanged

Note that the four graphs beginning at the strongest base pairs illustrate almost the entire trend of amino acid gain and loss. We have to allow the emergence of polar amino acids from the GG base pair to account for cystein (GG→UG) and serine (GG→AG) gain.

Two short graphs – AC→AA (the replacement of threonine by asparagine) and UC→UA (the replacement of serine by tyrosine) – do not contradict the assumed prohibition on changing the polarity, but their contribution to the explanation of the trend is insignificant. The strong base pair of serine is lost and not accrued. This approximation only allows a gain of the weak base pair (AG) for serine.

We have to use another short graph, AA←GA→UA, to illustrate the loss of glutamic acid (Glu). This loss generates more questions than answers. The GA→AA replacement must result in lysine enrichment but although lysine is a polar amino acid, it is oppositely charged and it is quite consistently lost. The alternative loss of glutamic acid causing emergence of new stop codons is also fatal, but a shorter-chain-length protein may be more functional than the protein with a radically changed charge in one of the groups. Glutamic acid may also be lost due to the replacement of the codon's last letter, which converts it to aspartic acid. Then, aspartic acid will most probably be converted to tyrosine. Note that the GA→UA graph does not change.

The gradual loss of the strong base pair for arginine (CG) becomes inevitably necessary as the only way to account for the gain of histidine and serine. Based on this scheme, one can account for the gain of the abovementioned five amino acids and the loss of the four. Thus, the trend of amino acid gain and loss can be related to the evolution of DNA, leaving aside the evolution of the genetic code. Yet, there is a less pronounced tendency that has not been accounted for within the framework of these assumptions, namely, a nearly consistent gain of arginine and loss of lysine. Moreover, if arginine becomes a loser, lysine becomes a gainer, as reported by Jordan and colleagues (2005). These two polar, positively charged, amino acids with the extremely strong and weak base pairs behave as if they are not affected by the general direction of mutations, leading to AT-enrichment. If AT-enrichment is the leading mechanism in the formation of the trend, lysine must be a gainer. It can substitute for arginine by a single replacement in the weak base pair; it can also replace glutamine, asparagine, and, probably, glutamic acid. Replacements, beginning at the strong base pair of arginine, must result in a gradual loss of not only arginine but the whole pool of amino acids with positively charged side chains. It seems that there must be an active mechanism of arginine enrichment. This may indicate that there is a mechanism of active replacement of lysine by arginine (AA→CG). Much less puzzling is exchange of the non-polar glycine for the polar arginine as a result of the GG→AG replacement; this leads to the gain of the weak base pair and the loss of the strong one for arginine, leaving the lysine issue unaccounted for.

Thus, I proposed a hypothesis that accounts for the trend of amino acid gain and loss; to substantiate the hypothesis I used the variant of the table of the genetic code proposed by Rumer. The hypothesis gave rise to several questions, and the answers to them could prove, disprove or specify the original statement of the hypothesis. 1. How do the codons with weak and strong base pairs that encode arginine behave? 2. How do the base pairs for serine behave? 3. Finally, where do lysine and glutamic acid go?

Part 2. In Part 1 I showed that the trend of amino acid gain and loss can be accounted for through the G→T and C→T replacements in DNA. It was also stated that explanation of the trend does not have to involve the term "evolution of the genetic code" in the sense of the order in which various amino acids were recruited into the genetic code. Below I disprove the assumption that gainers are among the amino acids incorporated last.

To account for the extant structure of the genetic code (all dialects), I can propose the hypothesis of the evolution of the current four-letter alphabet from an earlier, two-letter one. In this hypothesis, deamination of cytosine plays the key role. Remembering that DNA and cytosine methylation seem to be of quite recent origin, it would be correct to write C→U.

Cytosine deamination obviously causes just partial loss of complementarity. There can be two hydrogen bonds between guanine and uracil, at first glance making this pair similar to the adenine-thymine (uracil) pair.

Based on the assumption that the two most complex nucleotides, guanine and thymine, were before uracil and adenine, one can reconstruct some of the stages of the genetic code evolution.

Long chains consisting of guanine and cytosine could encode the first four amino acids: proline, glycine, alanine, and arginine. In all dialects of the genetic code, these amino acids are encoded by the same base pairs. If there were adaptors between mRNA and these amino acids, they were of a very simple structure.

Deamination led to gradual accumulation of uracil, which was initially read as cytosine.

Due to accumulation of a considerable amount of uracil the new base pairs – CU, UC, and GU – acquired meaning. The presence of similar amino acids that correspond to these strong base pairs in all dialects is indicative of evolutionary antiquity of the base pairs and adaptors between mRNA and amino acids. The adaptors must have evolutionarily originated from their precursors and the amino acids were close to the antecedent ones in their chemical properties. Thus, we get Pro→Ser, Pro→Leu and Ala→Val. Now we can revise Table 3 of Part 1, allowing accumulation of the strong base pair for serine.

At that stage, the UU base pair did not acquire meaning because it was rare: the amount of uracil was still low.

The UG base pair could have acquired meaning later than CU, UC, and GU as cysteine could hardly have originated from the codon and adaptor corresponding to arginine. The most probable source of cysteine in the genetic code table is modification of the adaptor for glycine as a result of the oxidative guanine damage. The emergence of thymine (uracil) caused by the oxidative guanine damage occurs in the presence of adenine.

Examination of the symmetry of the genetic code table seldom involves use of Rumer's canonical sequence: C>G>U>A. Much more frequently used is Crick's sequence: U>C>A>G (Crick, 1968). The two different ways to arrange nucleotides in sequences highlight different properties of the code. Rumer's sequence indicates the number of hydrogen bonds in complementary pairs and Crick's sequence – the indistinguishability of A and G (C and U). But then why is it always that C>G and U>A? Why do we compare complementary nucleotides? There is a simple answer to this question: because they were incorporated into the genetic code at different times. When there only were guanine and cytosine, it was senseless to assert that C>G. Uracil was incorporated into the code before adenine was, and this is the basis for stating that U>A. The procedure of new codons originating from old ones implies that this relationship also becomes valid for the initial nucleotides: C>G.

The uracil-guanine pair must have distorted the complementarity of the neighboring nucleotides. Selection resulted in the emergence of adenine as a better pair for uracil. The relatively late emergence of adenine is indicated by the fact that differences in dialects of the code are localized in the base pairs involving adenine, except the UG base pair. The AC, CA, AG, and GA base pairs must have acquired meaning before the base pairs consisting of only adenine and uracil just because the combination of two new letters was rare. This order of filling the cells of the table caused glutamic acid, which is a loser, to be incorporated at almost the same time as the accruing histidine and later than serine. There is no indication that cysteine could have emerged later than, say, lysine.

The possibility of filling new cells in the genetic code table by amino acids from different antecedent cells yielded various adaptors and weak base pairs.

Thus, the leading mechanism of the trend of amino acid gain and loss is the leading mechanism of the evolution of the genetic code. This accounts for the stability of the code towards point mutations and the final symmetry of the genetic code table.

Acknowledgement
The author would like to thank Krasova E. for her assistance in preparing this manuscript.